\documentclass[12pt]{article}

\usepackage{a4}
\usepackage{epsf}
\usepackage{amsmath}
\usepackage{subfigure}
\usepackage{graphicx}
\begin{document}

\title{Effect of hard processes on momentum correlations in $pp$ and $p\bar{p}$  collisions}

\author{Guy Pai\'c\\[1ex]
{\it Instituto de Ciencias Nucleares, UNAM, Mexico City, Mexico}\\[1.5ex]
Piotr Krzysztof Skowro\'nski\\[1ex]
{\it CERN, CH--1211 Geneva 23, Switzerland and}\\
{\it Warsaw University of Technology, Faculty of Physics,}\\
{\it ul.~Koszykowa 75, 00-662 Warsaw, Poland}\\[1.5ex]
}

\date{2 April 2005}

\maketitle

\begin{abstract}

The HBT radii extracted in $p\bar{p}$ and $pp$ collisions at SPS and Tevatron 
show a clear correlation with the charged particle rapidity density.
We propose to explain the correlation using a simple model where
the distance from the initial hard parton-parton scattering 
to the hadronization point depends on the energy of the partons emitted. 
Since the particle multiplicity is correlated with the mean energy 
of the partons produced we can explain the experimental observations without 
invoking scenarios that assume a thermal fireball. The model has been 
applied with success to the existing experimental data both in the magnitude 
and the intensity of the correlation. As well, the model has been extended to pp
collisions at the LHC energy of 14 TeV. The possibilities of a better insight 
into the string spatial development using 3D HBT analysis is discussed.

\end{abstract}


\section{Introduction}

The size of the source created in $pp$ and $p\bar{p}$ collisions,
as measured with momentum correlations, increases with the particle multiplicity 
(\cite{UA1hbt},\cite{E735hbt}).
The correlation of the extracted size with the rapidity density of the collisions,
from the HBT analysis, was sometimes described as  evidence for the existence of a 
"source" with a given size. Some alternative explanations have been given invoking 
long lived resonances and multiple parton interactions \cite{Buschbeck:2000rd}.

In the present work we are investigating whether the observed behavior may 
be understood in terms of more trivial explanations related to the details of the
hadronization of the partons leading to jets.
We know namely, that the point of hadronization of a jet and the point of the initial 
parton--parton hard scattering do not coincide. 
The distance between them is the so called hadronization length ($L_{hadr}$).

\begin{figure}
  \subfigure[]{\includegraphics[width=0.49\columnwidth] {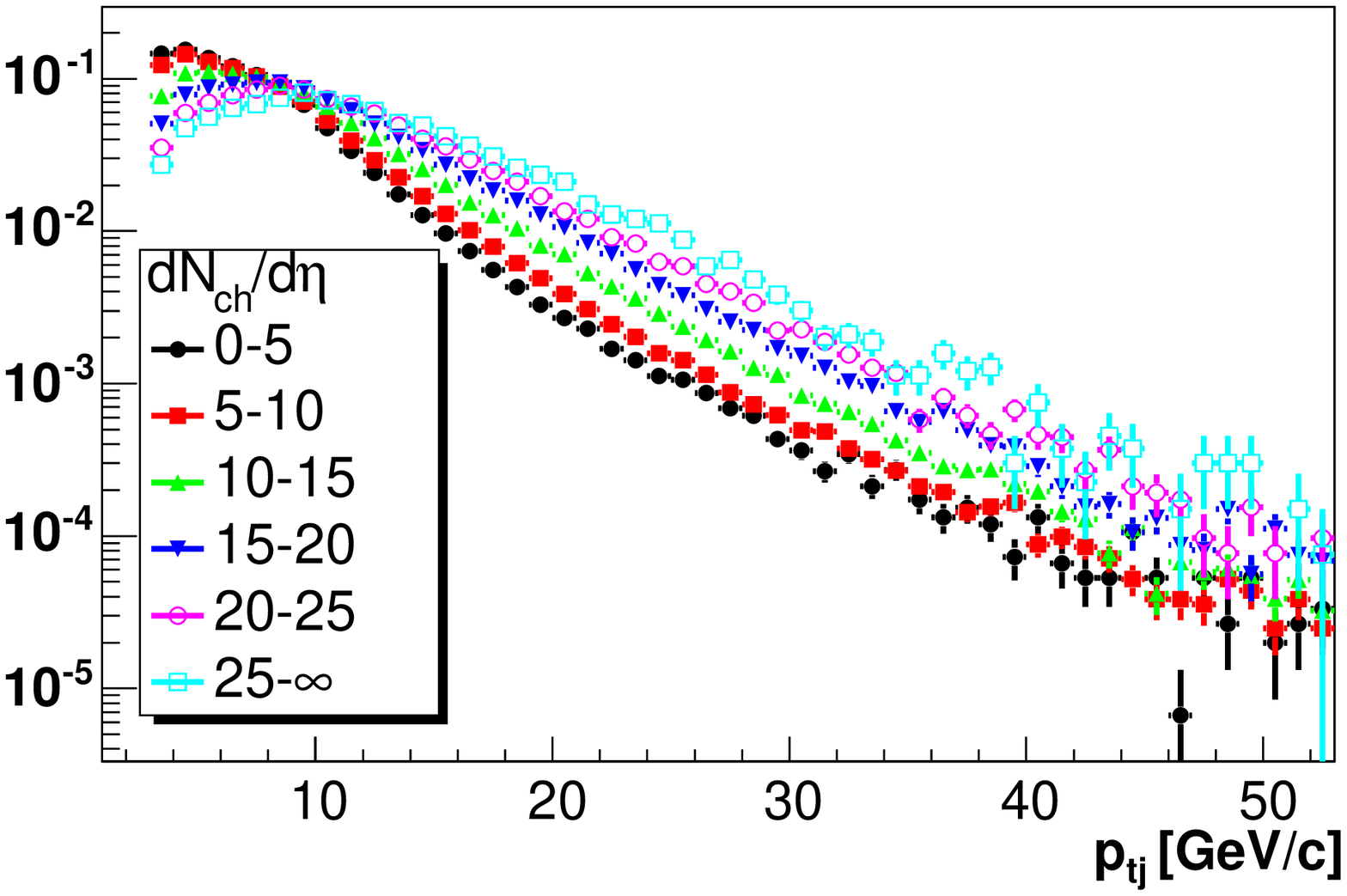}}
  \subfigure[]{\includegraphics[width=0.49\columnwidth] {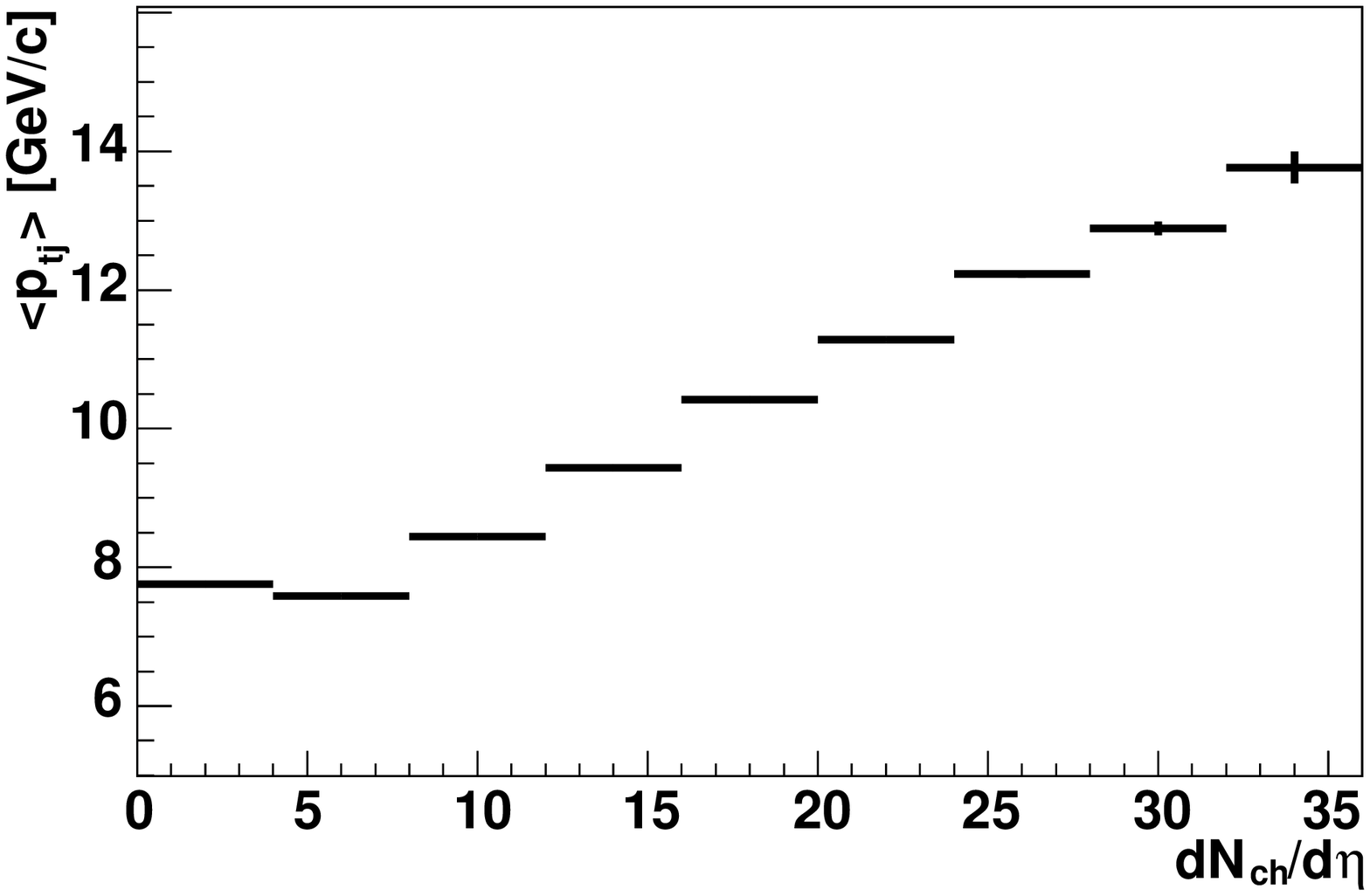} }
  \caption{a) Distribution of the transverse jet momentum, for different $dN_{ch}/d\eta$ ranges. 
              The cut at 3 GeV/c applied in the calculations is visible in the figure.
           b) Distribution of the mean transverse jet momentum ($p_{tj}$) 
              versus $dN_{ch}/d\eta$. 
           The results were obtained with Pythia at $\sqrt{s}=1.8$~TeV.}
  \label{cap:meanetvsnch}
\end{figure}

Numerical estimates for the time scale of hadronization vary significantly 
\cite{Wang},\cite{Dokshitzer},\cite{Kopeliovich}, 
but owing to the Lorentz boost to the laboratory frame, they are proportional to the energy,
$L_{hadr} \sim O(1) E_{t}$ (\cite{Wiedemann:2004wp}).
Hence, there is a dependence of $L_{hadr}$ on the energy due to 
the Lorentz boost. On the other hand the energy spectrum of the emitted jets depends on the 
charged particle multiplicity of the events as shown in Fig.\ref{cap:meanetvsnch}.
Hence if we assume that the hadronization occurs at different distances from 
the initial hard scatterings, depending on the energy of the jet, we can expect that 
this effect may simulate an extended hadronic source without invoking the presence of a 
thermalized source of hadrons. We have followed this line of thought in the present work.

\section{Simulation}
\label{simul}

The simulation comprises three steps:
\begin{enumerate}
\item The simulation of the particle momenta using a standard Pythia event generator. 
      Identification of jets and "underlying event".
\item Creation of a spatial distribution of particle origins according to our perception 
      of the hadronization process.
\item Implementation of the Bose-Einstein effect and creation of the correlation functions.
\end{enumerate}

\subsection{Event generation and jets identification}

With Pythia 6.24 \cite{Pythia} we simulate pp collisions with $\sqrt{s}=1.8$\ TeV. 
The PYCELL subroutine, that is part of the generator, is used to identify jets. 
We applied the following parameters:
\begin{itemize}
\item pseudo-rapidity range ($\eta$): from -2 to 2
\item number of pseudo-rapidity bins: 1200
\item number of bins in azimuthal angle: 1200  
\item threshold transverse energy of particles considered: 0
\item minimum transverse energy of particles that are used as jet seeds: $0.7$ GeV
\item minimum jet transverse energy: $3$ GeV
\item maximum jet radius $R=\sqrt{\Delta\phi^2+\Delta\eta^2}$: 1
\end{itemize}

All particles that do not belong to any jet are treated as an "underlying event".
Jet axis $\vec{p}_{j}$ - the direction along which the jet develops - is defined as
\begin{eqnarray}
  p_{tj}      &=& \sum_{i}{p_{ti}} \\
  \phi_{j}    &=& \frac{\sum_{i}{\phi_{i} p_{ti}}}{p_{tj}} \\
  \eta_{j}    &=& \frac{\sum_{i}{\eta_{i} p_{ti}}}{p_{tj}} \\
  \vec{p}_{j} &=& (p_{xj},p_{yj},p_{zj}) = p_{tj}(\cos\phi_{j},\sin\phi_{j},\sinh\eta_{j})
\end{eqnarray}
The sums run over all particles that make up a jet. $p_{ti}$, $\phi_{i}$ and $\eta_{i}$
are transverse momentum, azimuthal angle and pseudorapidity of the ith particle, respectively.

\subsection{The Source Models}
The events simulated with Pythia are then treated according to our model. 
Namely, the particles identified above as "underlying events" are given a spatial origin 
centered around the initial hard scattering point, while particles within a jet are given 
spatial coordinates of origin according to one of the models described below.

\paragraph{Tube}
It is assumed that the hadronization length ($l_{j}$, the distance from the
hard scattering) depends linearly on the initial parton energy that
we approximate by $p_{tj}$ (jet total transverse momentum). 
Thus, $l_{j}=f_{l}p_{tj}$, where  $f_{l}$ is a multiplicative factor that represents our
lack of theoretical insight into the process of hadronization.
For every parton the loci of hadronization along the jet axis ($x_{l}$) 
is randomized from Gaussian distributions with a mean equal to $l_{j}$ and
a $\sigma_{l}=l_{j}/3$, preventing negative values.  
In the transverse direction (with respect to the jet axis) the hadronization 
points are randomized so that the distance to the jet axis follows a Gaussian 
distribution with a variance equal to $\sigma_{t}$ and mean value of zero.

\paragraph{Dynamic width}

The distribution along the jet axis is the same as above while
the transverse width $\sigma_{t}$ depends linearly on the jet transverse energy and 
moreover, it is a function of the position along the jet axis (see Fig.\ref{cap:kine}) 
so the distribution of hadronization points is:

\begin{equation}
\label{bgb}
\sigma_{t}(x_{l},p_{tj}) =
\begin{cases}
\sigma_{t}^{\rm max}\exp\frac{-(l_{j}-x_{l})^2}{w}
& \mbox{if}\quad \sigma_{t}^{\rm max}>\sigma_{t}^{\rm min}
\\
\sigma_{t}^{\rm min} 
& \mbox{if}\quad \sigma_{t}^{\rm max} \le \sigma_{t}^{\rm min}
\end{cases}\, ,
\end{equation}
where $\sigma_{t}^{\rm max} = f_{t}p_{tj}$, $\sigma_{t}^{\rm min} = 0.5$ fm and
$w = \frac{{l_{j}}^2}{\ln{2\sigma_{t}^{\rm max}}}$ 
($w$ is chosen so $\sigma_{t}(x_{l}=0)$ and $\sigma_{t}(x_{l}=2l_{j})$ are equal to 
$\sigma_{t}^{\rm min}$).

\paragraph{}

For both kinds of geometries the hadronization time is equal to $x_{l}$.

The positions of the "underlying event" particles are
randomized from a single Gaussian distribution with variance $\sigma_{b}$.
Their emission time is always equal to 0.

\subsection{Simulation of BE correlations}
Since the generator does not provide for Bose-Einstein correlations 
they have to be introduced. In our simulation we introduce them using the 
weighting algorithm due to Lednick\'y \cite{Weights}. 
It is applied during the construction of the correlation functions.
Each particle pair $(i,j)$ is weighted with a probability $\rho_{ij}$
  \begin{eqnarray}
    \label{2a}
    C(Q,K) &=&
    \textstyle{\frac{1}{N(Q,K)}}
    \sum_{(i,j)}\, \rho_{ij}\, , \\
    \label{2b}
    \rho_{ij} &=& 1 + \cos( ({\bf p}_i-{\bf p}_j)\cdot ({\bf x}_i-{\bf x}_j))\, .
  \end{eqnarray}
where
$N(Q,K)$ is the number of pairs in a given bin, 
$Q = p_i-p_j$ and $K = (p_i+p_j)/2$, 
${\bf x}_i$ and ${\bf p}_i$ are the 4-vectors of the hadronization points
and momentum in the pair rest frame, respectively.
The probability $\rho_{ij}$ coincides with the formal Born probability 
density ${\Psi}^*\Psi$ of the Bose-Einstein symmetrized 2-particle plane wave. 
%
  \begin{eqnarray}
    \label{3a}
    \rho_{ij} &=& \Psi^*({\bf x}_i,{\bf x}_j,{\bf p}_i,{\bf p}_j)\, 
                  \Psi  ({\bf x}_i,{\bf x}_j,{\bf p}_i,{\bf p}_j)\, ,\\
    \label{3b}
    \Psi({\bf p}_i,{\bf p}_j,{\bf p}_i,{\bf p}_j) &=& {1\over \sqrt{2}}
    {\left({ {\rm e}^{i{\bf p}_i {\bf x}_i + i{\bf p}_j {\bf x}_j}
           + {\rm e}^{i{\bf p}_j {\bf x}_i + i{\bf p}_i {\bf x}_j}
          }\right)}\, .
  \end{eqnarray}

\subsection{Correlation Functions}

The correlation functions are calculated for particles with $|\eta|<1$ and $p_t > 0.1$ GeV,
while no such a constraint is imposed in the jet finding procedure. 
We extract the correlation functions with two types of parameterizations.
\begin{itemize}
\item 
$C(Q_{t},Q_{0})=1+\lambda
\exp(
\frac{-1}{\hbar^2c^2}
({Q_{t}}^{\rm 2}{R_{t}}^{\rm 2}+
{Q_{0}}^{\rm 2}{\tau}^{\rm 2}))$
\item 
$C(Q_{out},Q_{side},Q_{long})=1+\lambda\exp(\frac{-1}{\hbar^2c^2}(
                                             Q_{out}^{2}R_{out}^{2}\
                                            +Q_{side}^{2} R_{side}^{2}\
                                            +Q_{long}^{2}R_{long}^{2})) $
\end{itemize}
where:
\begin{itemize}

\item $Q_{t}$: is the component of the three-momentum difference 
               perpendicular to the three-momentum sum 

\item $Q_{0}$: is the difference of energies

\item $Q_{long}$, $Q_{out}$ and $Q_{side}$: are the components of 
3-momentum difference vector
in the Longitudinally Co-Moving System (LCMS). $Q_{long}$ is parallel to beam,
$Q_{side}$ is perpendicular to beam and total pair momentum, 
and $Q_{out}$ is perpendicular to $Q_{long}$ and $Q_{side}$ (Fig.\ref{cap:kine}).

\item $R$'s: corresponding radii
\item $\tau$: dispersion (radius) in the time domain 

\begin{figure}[htb]
\vspace*{0.1cm}
\begin{center}
  \includegraphics[width=.37\textwidth]{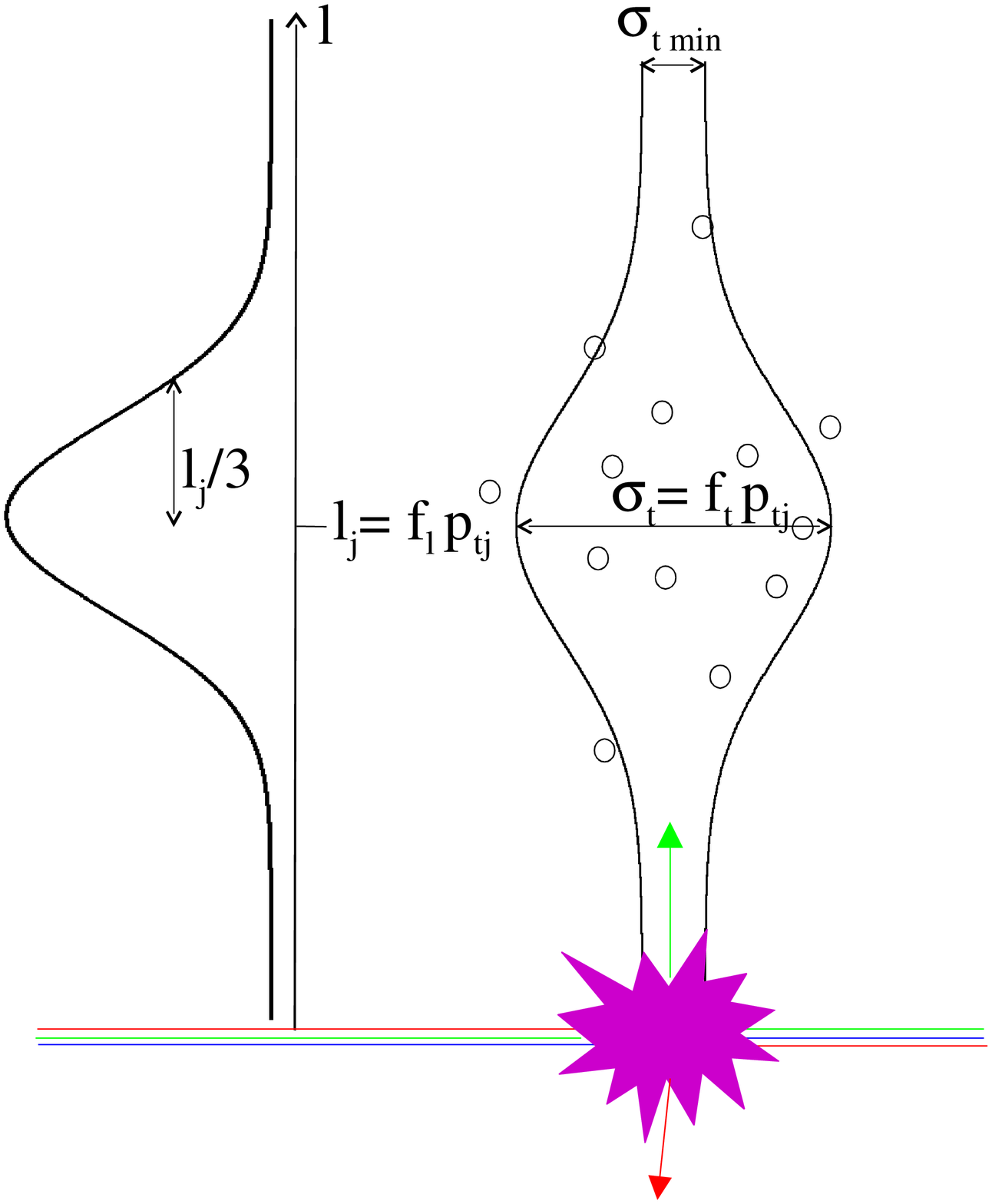}
  \includegraphics[width=.61\textwidth]{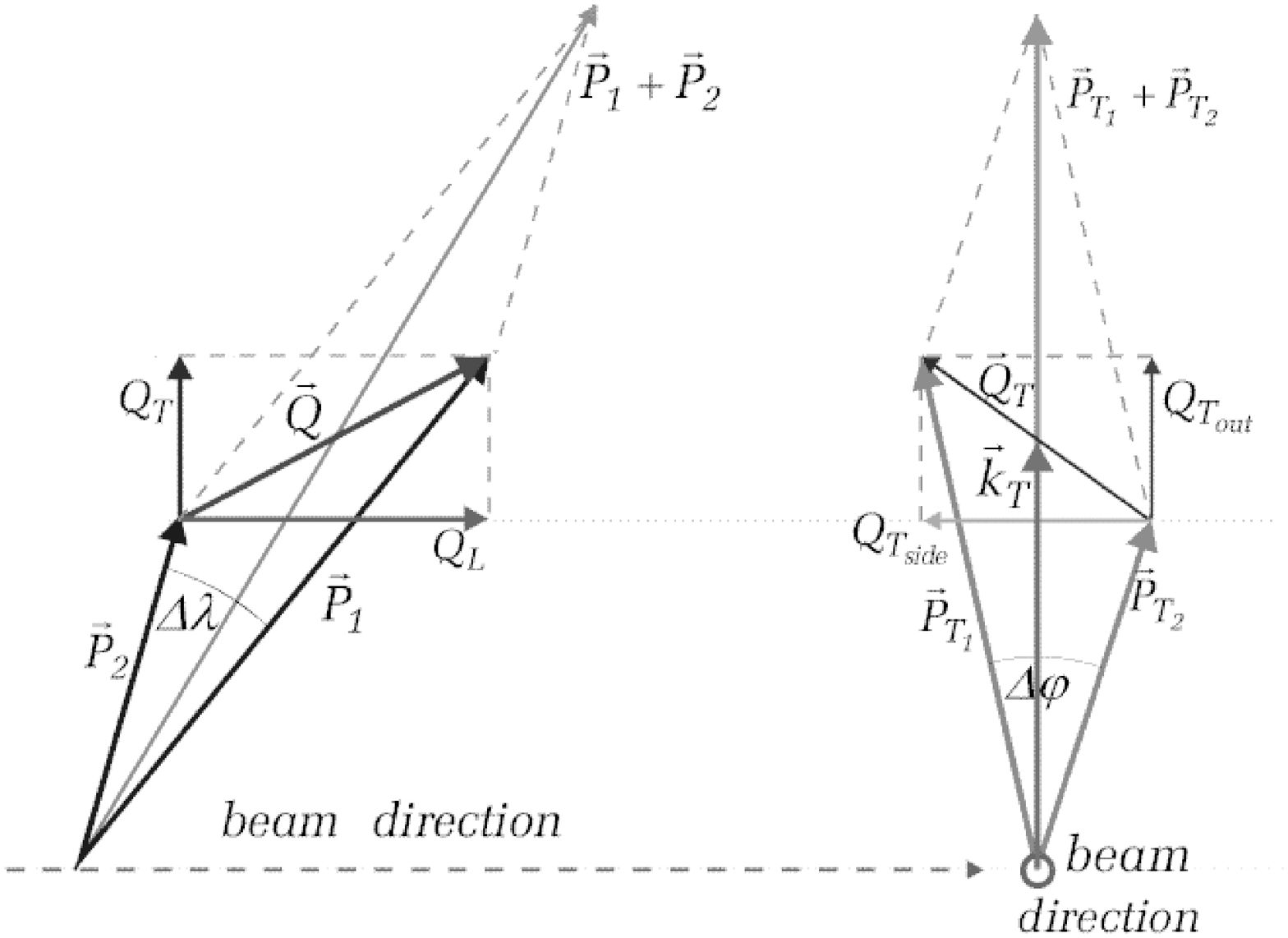}
\end{center}
\vspace*{-0.3cm}
\caption{a) Schema of the the "dynamic width" jet geometry model. 
            The spread in transverse direction with respect to the jet axis 
            depends on the position along jet and its magnitude depends 
            on the jet energy.
         b) Definition of $Q_{out}$, $Q_{side}$ and $Q_{long}$.}
\label{cap:kine}
\end{figure}

\end{itemize}

When speaking of the correlation function C($Q_{t}$) we mean the projection of 
$C(Q_{t},Q_{0})$ on the $Q_{t}$ axis for $|Q_{0}|<200$ MeV. 
By a double Gaussian fit we mean a 1D fit with the following form of a correlator
\begin{equation}
 C(Q)=1+\lambda_{1} e^{ - \left( \frac{QR_{1}}{\hbar c}\right)^2} 
       +\lambda_{2} e^{ - \left( \frac{QR_{2}}{\hbar c}\right)^2} 
 \label{eq:dblgaus}
\end{equation}

\section{Results}

The application of the model described above, in agreement with the intuition, 
shows that the correlation function indeed changes its shape with increasing multiplicity. 
Using the experimental data we have attempted to adjust the parameters of the model.
We have fixed $\sigma_{t}=0.5$ fm what corresponds to the typical hadronic size. 
We found that $\sigma_{b}=0.4$ fm   reproduces the experimental results
at the lowest multiplicities (E735 has measured $R_t=0.62$ fm at $<dN_{ch}/d\eta>=6.75$). 
In the frame of our model it implies a non-negligible 
contribution of hard processes in total particle production even at low multiplicities.
This observation is in an agreement with other observations at Tevatron \cite{Moggi:2004fu}.

However, we were not able to reach compatibility with the experimental values of $R_{t}$
at high multiplicities. The increase of the $f_{l}$ parameter causes a decrease of the 
intercept parameter, while the  shape of the correlation function stays approximately 
unchanged. In fact, the width of the peak, thus $R_t$ as given
by a Gaussian fit even decreases with increasing $f_{l}$.

Using the {\it out-side-long} (OSL) parametrization we have observed that for this model $R_{out}$ 
grows with $f_{l}$, while $R_{side}$ stays approximately unchanged. This finding matches the intuitive representation. Therefore the applied geometry with a constant 
  $\sigma_{t}$ limits the growth of $R_{t}$. We deduce 
 that $R_{side}$ must also increase with the jet energy. This led us to 
the "dynamic width" jet geometry.

\begin{figure}
\begin{center}
  \includegraphics[width=0.66\columnwidth] {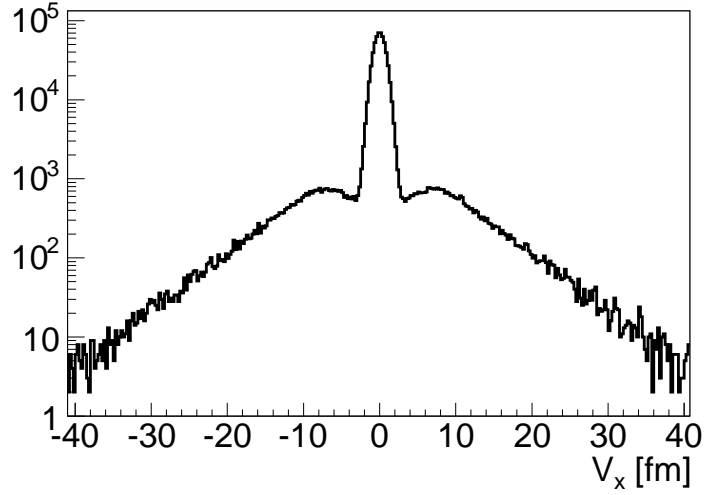}
  \caption{Cross-section through the 2D distribution of the hadronization points in the plane 
           perpendicular to the beam. The difference in height between contributions 
           from the "underlying event" (the peak for values around 0) and 
           jets (shoulders) is a phase space effect. 
           In fact, the majority  of particles originate from jets.
           The dynamic width jet geometry with $f_l=1.0$, $f_t=0.6$ 
           and $<dN_{ch}/d\eta>=12.4$.}
  \label{fl1.0ft0.6m3Vtxs}
\end{center}           
\end{figure} 

\begin{figure}
\begin{center}
  \subfigure[]{\includegraphics[width=0.49\columnwidth] {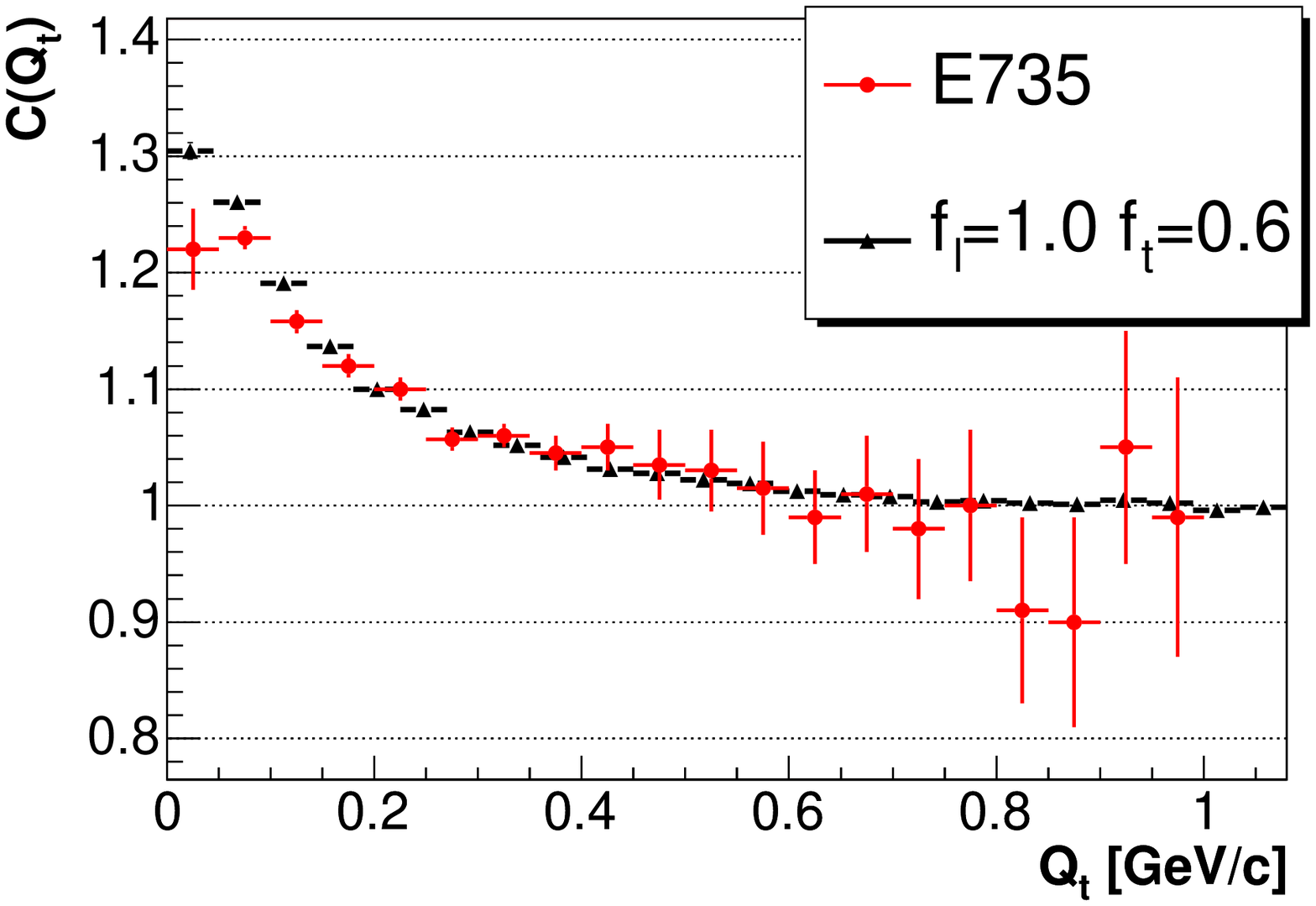}}
  \subfigure[]{\includegraphics[width=0.49\columnwidth] {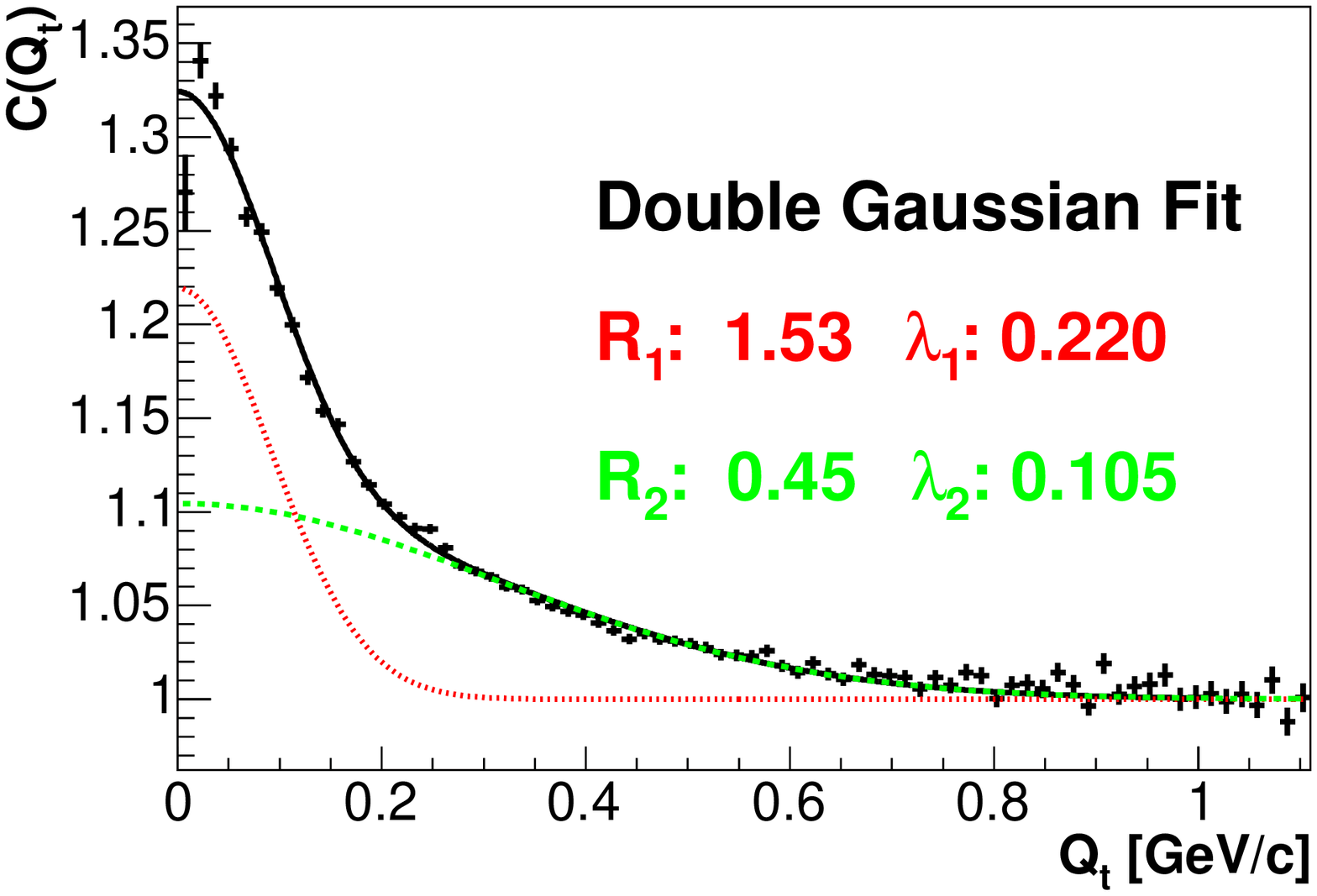}}
  
  \caption{ a) The $Q_t$ correlation function from our model compared with the correlation 
               function extracted from \cite{E735hbt} ($<dN_{ch}/d\eta>=12.5$) and 
            b) double Gaussian fit to it.
               The dynamic width jet geometry with $f_l=1.0$, $f_t=0.6$ and 
               $<dN_{ch}/d\eta>=12.4$.}
  \label{16m3}
\end{center}           
\end{figure}

\begin{figure}
  \subfigure[]{\includegraphics[width=0.49\columnwidth] {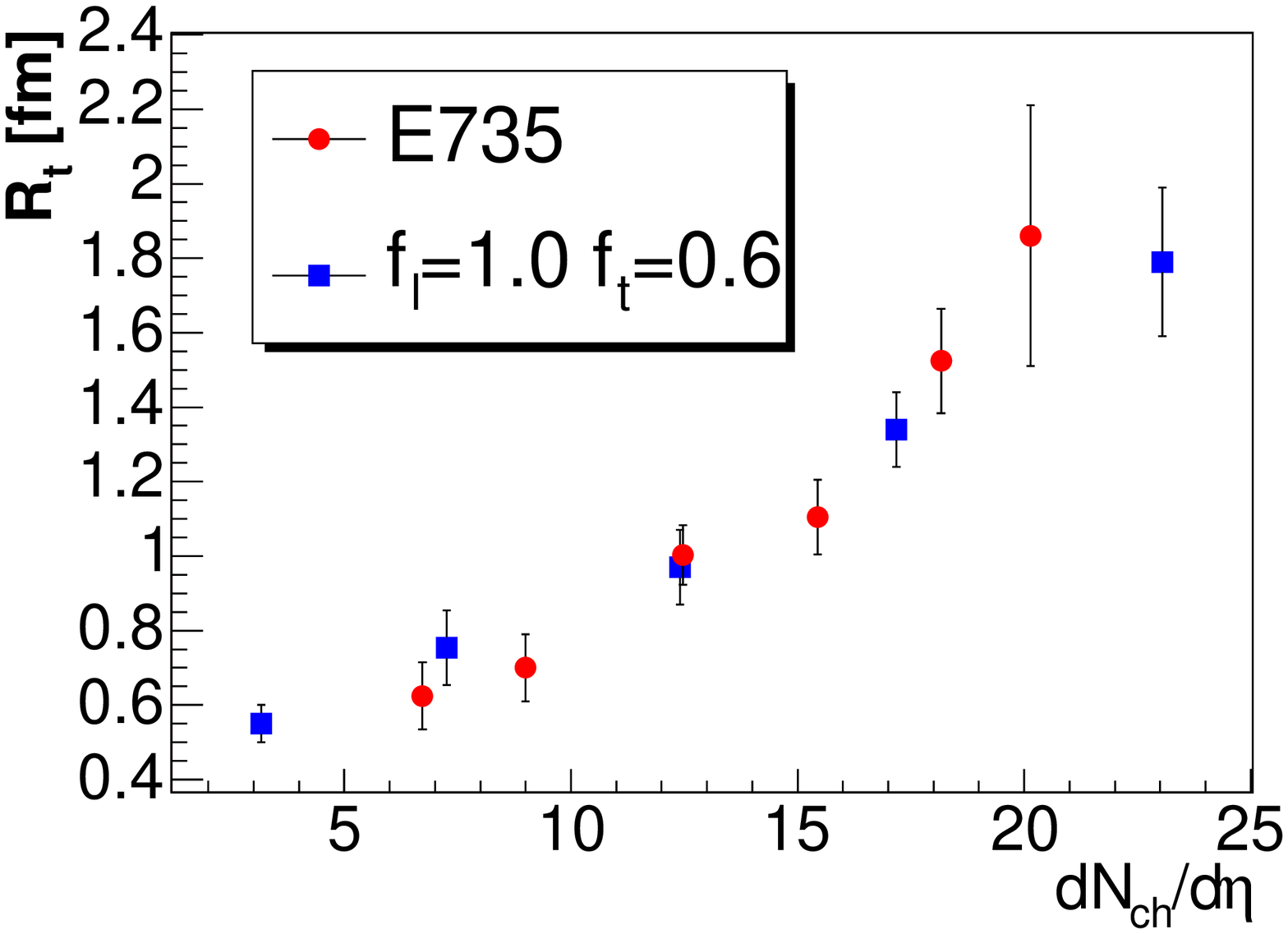} }
  \subfigure[]{\includegraphics[width=0.49\columnwidth] {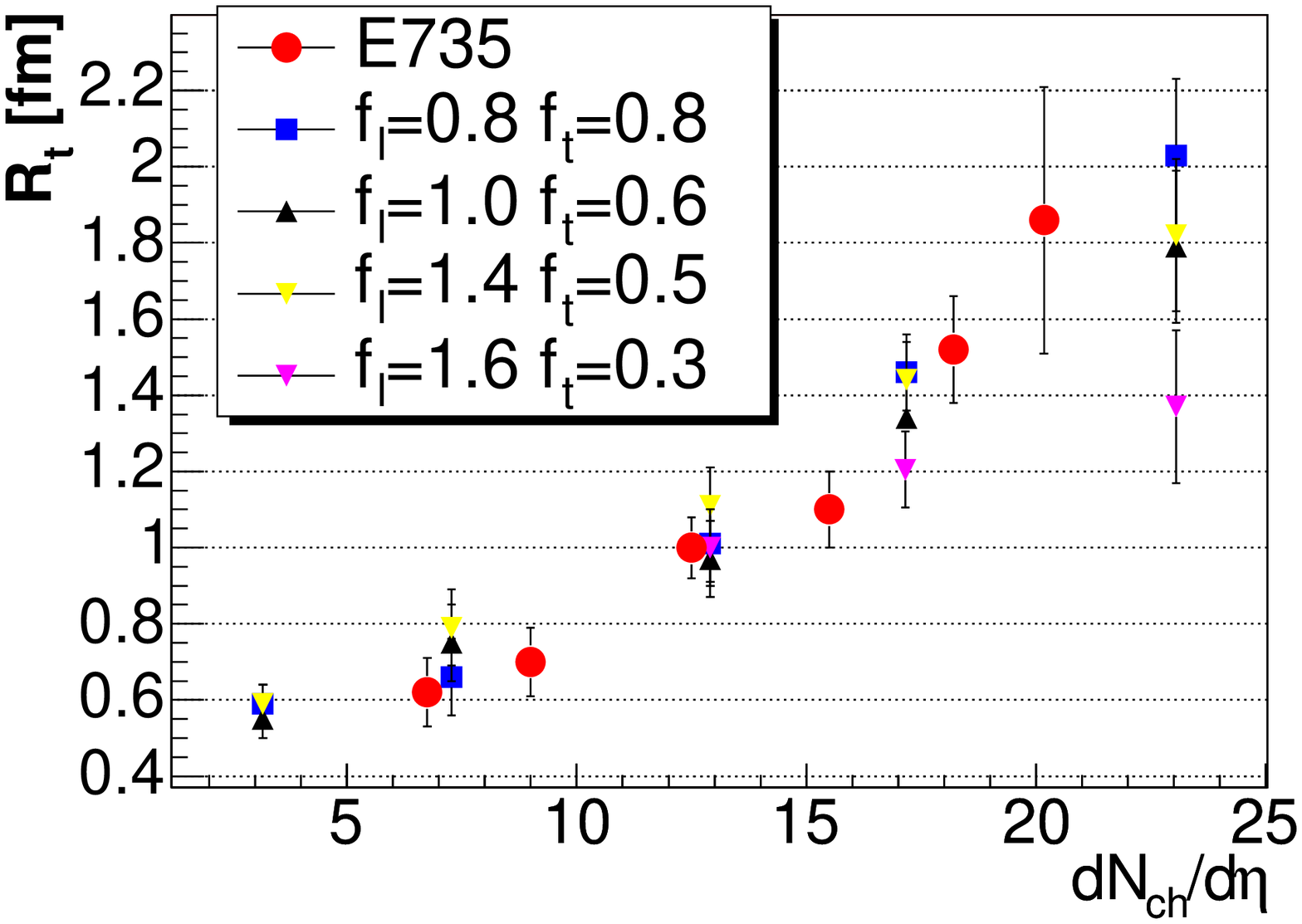}}
  \caption{ Results of our model compared to E735 result. 
            The left hand plot shows the best fit to the data obtained with 
            $f_l=1$ and $f_t=0.6$ while the right hand plot demonstrates the 
            variation of the results with varying the $f$ values.}
  \label{result}
\end{figure}

Using that model, we have found that we are able to reproduce the experimental
results with $f_{l}=1.0$ and $f_{t}=0.6$, see Fig.\ref{result}a and \ref{16m3}.
This is not a unique pair of parameters that gives a good agreement with E735 result. 
Within some range we can decrease $f_{l}$ and find such a value of $f_{t}$ such 
that we still reproduce the experimental result (Fig.\ref{result}b).

\begin{figure}
 \begin{center}
  \subfigure[]{\includegraphics[width=0.49\columnwidth] {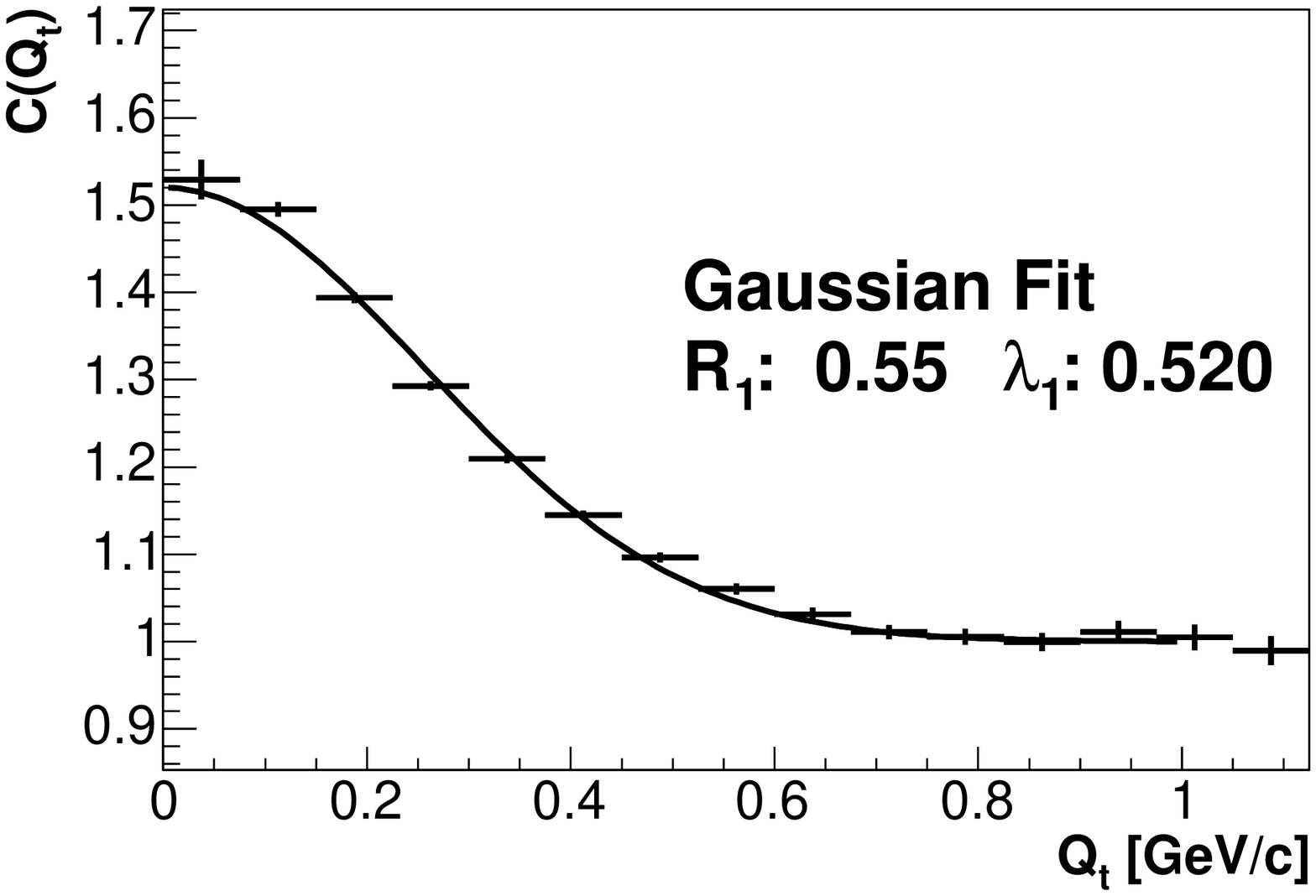}}
  \subfigure[]{\includegraphics[width=0.49\columnwidth] {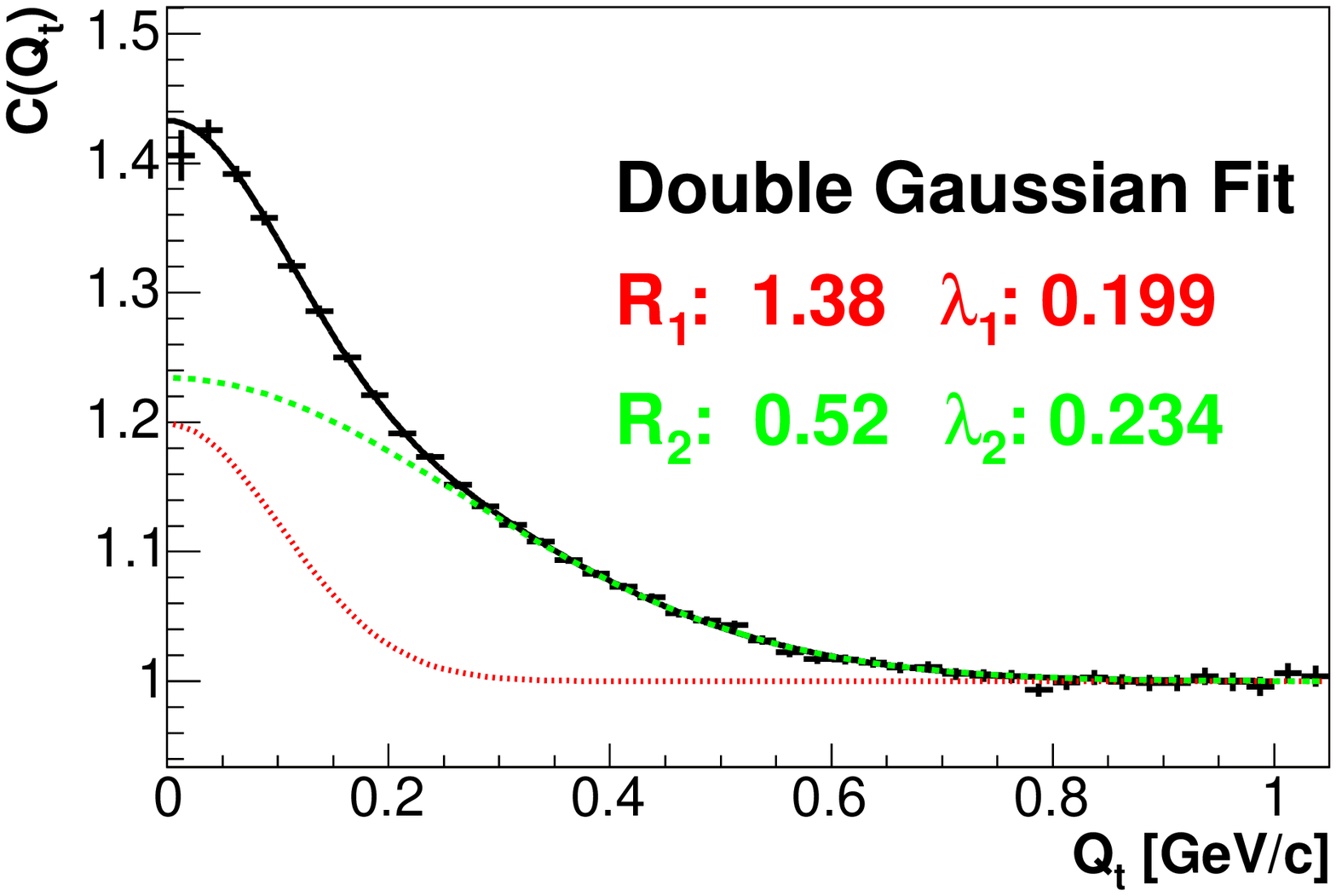}}

  \subfigure[]{\includegraphics[width=0.49\columnwidth] {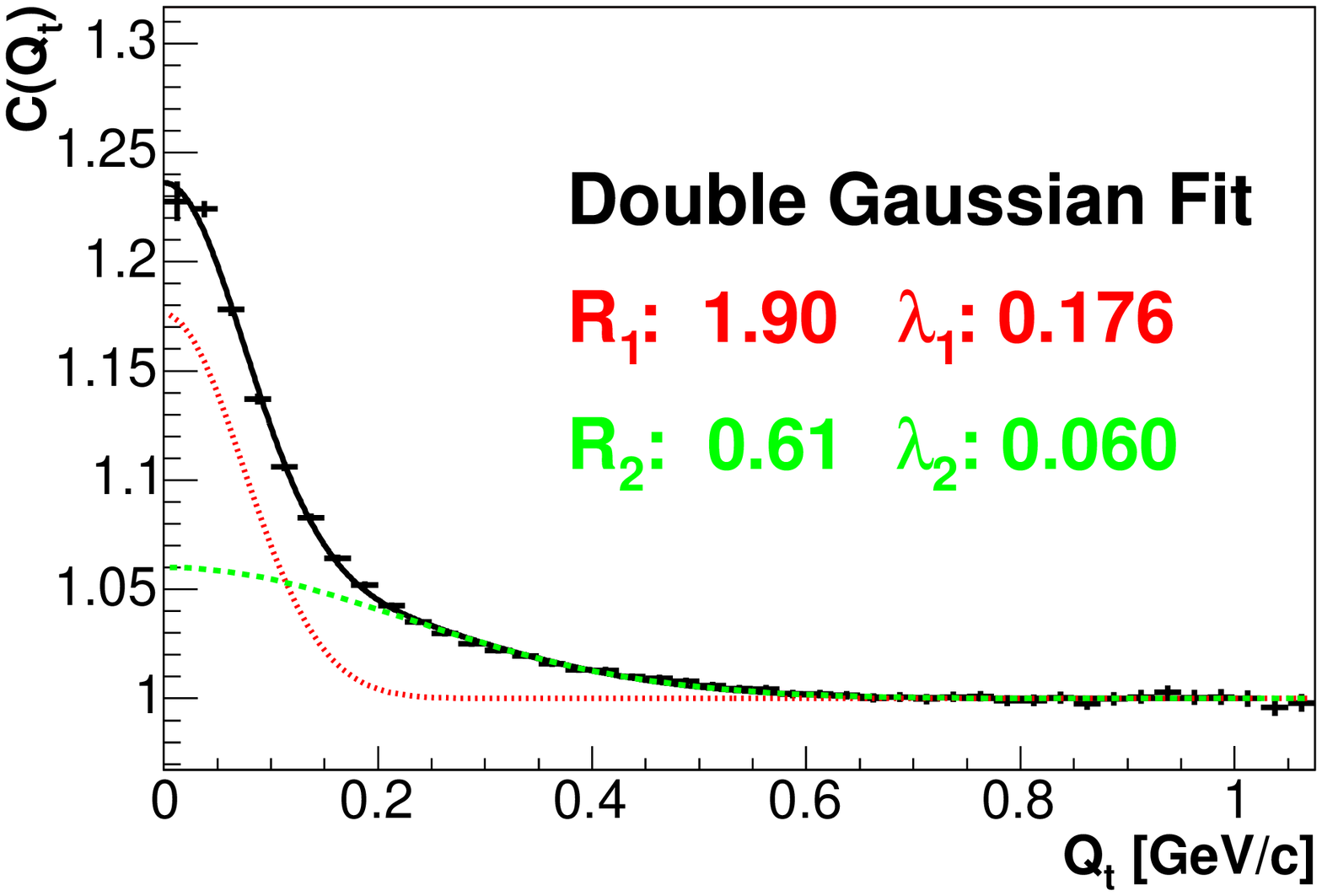}}
  \subfigure[]{\includegraphics[width=0.49\columnwidth] {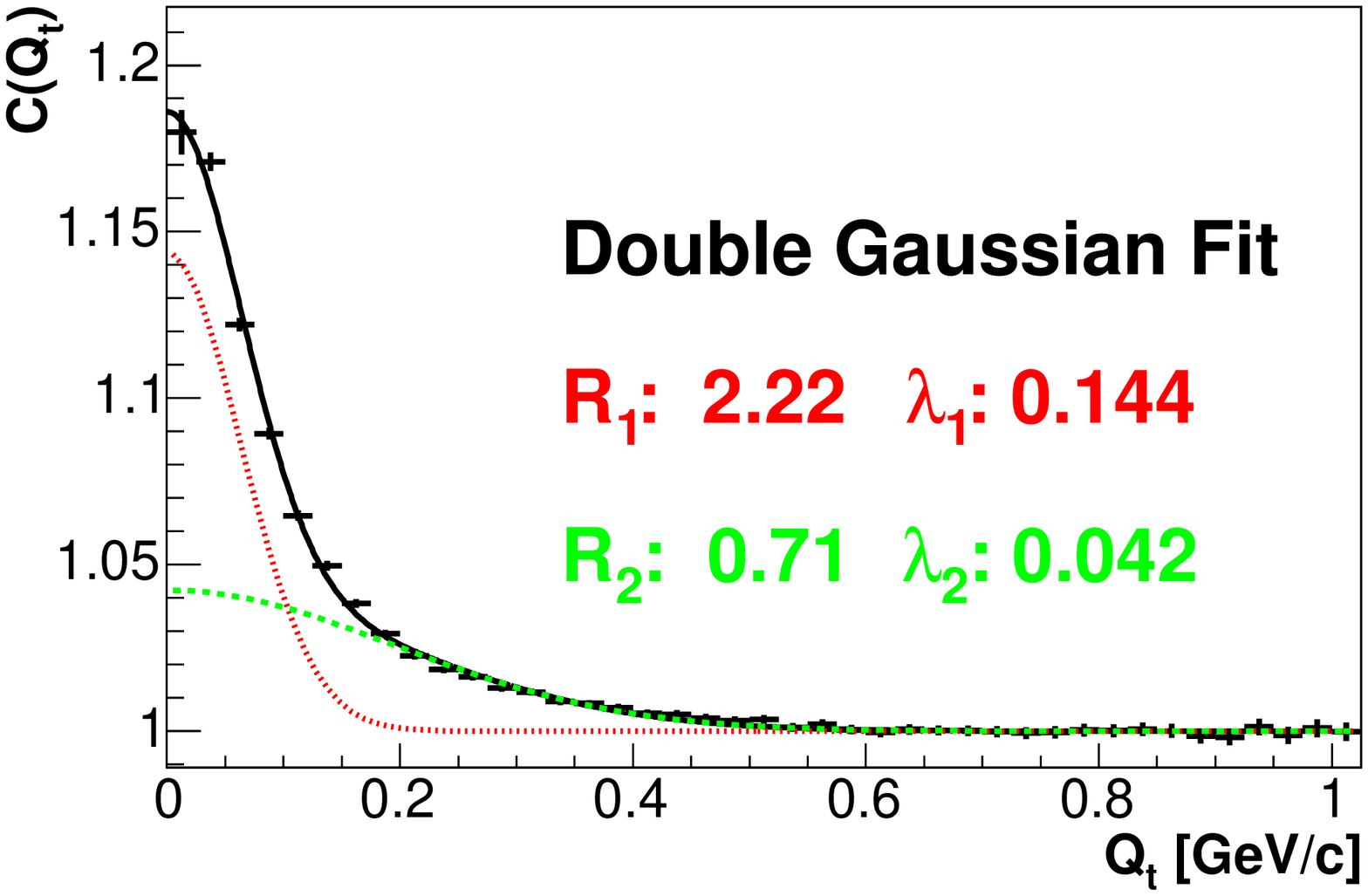}}

  \caption{ $Q_t$ correlation functions for the case of the dynamic width 
             geometry with $f_l=1.0$ and $f_t=0.6$ 
	     a) $dN_{ch}/d\eta =  3.2$,
	     b) $dN_{ch}/d\eta =  7.3$,
	     c) $dN_{ch}/d\eta = 17.2$,
	     d) $dN_{ch}/d\eta = 23.0$}
  \label{fl1.0ft0.6}
 \end{center}
\end{figure}

We believe that the precision on the determination of f values could be improved 
if results of a 3D HBT analysis were available.
Namely, the dependence of $R_{out}$ and $R_{side}$ on event multiplicity is required. 
We have found that $R_{side}$ increases together with $f_{t}$, and $R_{out}$ 
together with $f_{l}$ (see Table.\ref{tab:table1}). It means that we can estimate 
the size of the jet fragmentation volume using the three dimensional 
correlation analysis.

\begin{table}
\begin{center}
\begin{tabular}{|c|c||c|c|c|c||c|c|c|c||c|c|c|c|}
  \hline
  $f_l$&
  $f_t$&
  $R_{o1}$&
  $\lambda_{1}$&
  $R_{o2}$&
  $\lambda_{2}$&
  $R_{s1}$&
  $\lambda_{1}$&
  $R_{s2}$&
  $\lambda_{2}$&
  $R_{l1}$&
  $\lambda_{1}$&
  $R_{l2}$&
  $\lambda_{2}$
  \tabularnewline
        &
        &
  [fm]&
  &
  [fm]&
  &
  [fm]&
  &
  [fm]&
  &
  [fm]&
  &
  [fm]&
  \tabularnewline\hline
  0.2&
  0.2&
  0.82&
  0.55&
  -&
  -&
  1.33&
  0.61&
  -&
  -&
  1.28&
  0.58&
  -&
  -

  \tabularnewline\hline
  0.25&
  0.25&
  0.92&
  0.50&
  -&
  -&
  1.51&
  0.55&
  -&
  -&
  1.44&

  0.53&
  -&
  -

  \tabularnewline\hline
  0.3&
  0.3&
  1.02&
  0.47&
  -&
  -&
  1.65&
  0.52&
  -&
  -&
  1.56&
  0.50&
  -&
  -

  \tabularnewline\hline
  0.4&
  0.4&
  2.12&
  0.23&
  0.84&
  0.23&
  2.02&
  0.44&
  0.51&
  0.03&
  2.42&
  0.40&
  0.75&
  0.09

  \tabularnewline\hline
  0.5&
  0.5&
  2.30&
  0.24&
  0.79&
  0.19&
  2.22&
  0.42&
  0.40&
  0.03&
  2.47&
  0.37&
  0.61&
  0.08

  \tabularnewline\hline
  0.6&
  0.6&
  2.70&
  0.19&
  0.87&
  0.17&
  2.77&
  0.31&
  1.01&
  0.08&
  3.11&
  0.28&
  1.04&
  0.12

  \tabularnewline\hline
  0.8&
  0.8&
  2.88&
  0.21&
  0.67&
  0.11&
  2.84&
  0.28&
  0.69&
  0.05&
  3.41&
  0.25&
  0.84&
  0.09

  \tabularnewline\hline
  1.0&
  0.6&
  3.03&
  0.25&
  0.66&

  0.11&
  2.75&
  0.28&
  0.67&
  0.06&
  3.00&
  0.27&
  0.78&
  0.09

  \tabularnewline\hline
  1.4&
  0.5&
  3.52&
  0.25&
  0.64&
  0.09&
  2.55&
  0.26&
  0.75&
  0.06&
  2.89&
  0.23&
  0.86&
  0.09

  \tabularnewline\hline

\end{tabular}
\end{center}
  \caption{Dependence of $OSL$ radii on $f_l$ and $f_l$ for the $<dN_{ch}/d\eta>=23.0$.
          The fits were made imposing a double Gaussian  (Eq. \ref{eq:dblgaus}) 
          on 1D projections, taking two other components $>50$ MeV. 
          In the rows where $R_{2}$ is not specified
          fits did not converge and a single Gaussian is used instead.
          }
  \label{tab:table1}
\end{table}

In our model the "underlying event" (UE) is somewhat overestimated since we have 
imposed the cut on the jets of less than 3 GeV and this part has been added to the UE.
We have examined how our results change if we reduce UE by removing randomly 
50\% of particles not assigned to jets.
We have found that the obtained radii stay unchanged within 10\%.

\begin{figure}
  \begin{center}
    \includegraphics[width=0.66\columnwidth]{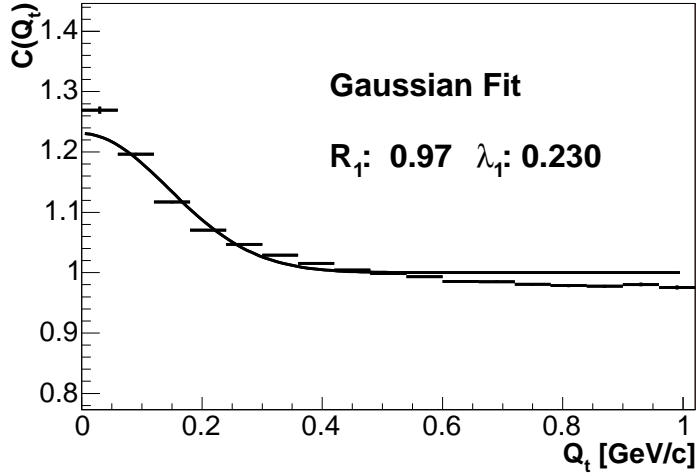}
    \caption{The same as in Fig.\ref{16m3}, normalized at $Q_{t} \sim 500$ MeV 
             and fitted with the single Gaussian form of the correlator.}
    \label{sglgaus}
   \end{center}
\end{figure}

From Fig.\ref{sglgaus} we see that the correlation functions extend up to large
Q's ($0.4 - 1.0\ GeV$), similarly to the ones obtained by E735. The slope of the ``tail''
depends on the multiplicity. The simulated correlation function cannot be well represented 
by a single Gaussian as expected from the distribution of particle hadronization 
points shown in Fig.\ref{fl1.0ft0.6m3Vtxs}. As it can be seen e.g. in Fig.\ref{fl1.0ft0.6}c,
a better fit of the correlation function is obtained using a double Gaussian, 
albeit it is not yet probably the exact form of a correlator for this kind of source.
The two radii may be understood in term of the smaller one representing 
the correlations among the particles from the "underlying event" and the larger 
one representing the correlations of jet particles with the ones of the 
"underlying event", although the interplay of the different factors make such a 
representation only partially true. 

It is important to mention that the extracted radii are much smaller than the extent of 
the source due to the fact that the particles from the "underlying event" are traveling 
while the jet did not yet hadronize! 
Similarly, particles hadronizing first within 
a jet also moves together with not yet hadronized partons. The same argumentation
also explains the weak dependence of $R_t$ on $f_l$ in the ``tube geometry''.

Fitting such a correlation function with a single Gaussian - as was done in E735 - 
brings large uncertainties {(Fig.\ref{sglgaus})}, because the obtained result is very 
sensitive to the normalization chosen (at which point correlation function crosses 1).
To be able to compare with the experimental results 
\cite{UA1hbt}\cite{E735hbt} we have nevertheless used the single 
Gaussian fit. The results of the double Gaussian fits are shown in 
{Fig. \ref{16m3} and \ref{fl1.0ft0.6}}.

Finally we have calculated, using the parameters extracted for the Tevatron data, 
the expected correlation of radii with charged particle multiplicities for the 
maximum LHC energy of 14 TeV.
In Table \ref{tab:table2} we present results of our model obtained at LHC energies.

\begin{table}
\begin{center}
\begin{tabular}{|c|c|c|c|c|}
  \hline
  $dN_{ch}/d\eta$&
  $R_{t1}$&
  $\lambda_{1}$&
  $R_{t2}$&
  $\lambda_{2}$
  \tabularnewline
        &
  [fm]&
  &
  [fm]&

  \tabularnewline\hline
  $3.4$&
  $1.15$&
  $0.15$&
  $0.50$&
  $0.41$

  \tabularnewline\hline
  $7.6$&
  $1.38$&
  $0.20$&
  $0.52$&
  $0.21$

  \tabularnewline\hline
  $12.5$&
  $1.70$&
  $0.18 $&
  $0.58 $&
  $0.09$

  \tabularnewline\hline
  $17.4 $&
  $1.95 $&
  $0.15 $&
  $0.65 $&
  $0.05$

  \tabularnewline\hline
  $22.4$&
  $2.24$&
  $0.12$&
  $0.73$&
  $0.03$

  \tabularnewline\hline
  $27.4 $&
  $2.91$&
  $0.10 $&
  $0.95$&
  $0.04 $

  \tabularnewline\hline
  $37.0$&
  $3.33$&
  $0.07$&
  $1.15$&
  $0.02$

  \tabularnewline\hline
\end{tabular}
\end{center}
  \caption{$R_t$ radii dependence on $dN_{ch}/d\eta$ at $\sqrt{s}=14$ TeV, 
           $f_l=1.0$ and $f_t=0.6$. The correlation functions were fitted with 
	   double Gaussian form of correlator (see Eq. \ref{eq:dblgaus}). }
  \label{tab:table2}
\end{table}


\section{Conclusions}
Using a simple approach which introduces a dependence of the distance of the 
mean hadronization points of a parton on its energy we have been able to 
reproduce very satisfactorily both the dependence of the radii, and the trend of 
the correlation strength lambda with the rapidity 
density in pp collisions. The present results 
indicate that there is a possibility of an alternative interpretation 
of the results to those presented in \cite{E735hbt} and \cite{Gutay:2003xj}
where the obtained radii are interpreted as evidence for the observation of 
deconfined matter in pp collisions,

On the other hand the model a posteriori justifies the hadronization 
scenario envisaged because the free parameter $f$ has been found close to unity 
for the range of multiplicities analyzed. 
We believe that the LHC with its wider range of multiplicities in pp collisions 
offers interesting possibilities to test our model. We have therefore presented 
here the expected variation of the radii in function of charged particle 
multiplicities at the LHC. Similarly, the effect of the parton hadronization 
should be taken into account in the analysis of  HBT radii in heavy-ion collisions
\cite{jetshbt}.


\section{Acknowledgments}

We would like to acknowledge the very useful discussions with {U. Wiedemann} 
and {J. Pluta}.


\end{document}